\documentclass[a4paper,11pt]{article}
\usepackage{epsf,multicol,ifthen}
\usepackage[dvips]{graphicx}
\newcommand{\bec}{\begin{center}}
\newcommand{\ec}{\end{center}}
\usepackage{cite}
\usepackage{hyperref}
\title{ Studying the behavior of the strong interaction constant at different energy scales}

\author{T. Obikhod\thanks{E-mail: obikhod@kinr.kiev.ua}\\
\small\emph{Institute for Nuclear Research, National Academy of Science of Ukraine} \\
\small\emph{47, prosp. Nauki, Kiev, 03680, Ukraine}}

\date{\small\today}

\begin{document}

\maketitle

\abstract{The character of the behavior of the RGE constant on a certain energy interval is studied and polynomial interpolation is carried out to obtain an analytic function describing the RGE constant in QCD. A graphical description of the nature of the dependence of the beta function and the interaction constant is carried out and it is shown that a good agreement with the theoretical predictions of the asymptotic freedom only in the energy region of 3800-4000 GeV. }

\section{Introduction}
\label{sec:intro}
The world of elementary particles is built from two components: quantum mechanics and special theory of relativity, united in relativistic quantum mechanics. The combination of field theory and relativistic quantum mechanics gives a quantum field theory. The quantum theory of the field is used to describe quantum electrodynamics, quantum chromodynamics operating with quantum fields, which are described by the Lagranjian minimization equation of interaction between particles associated with fields. The interaction between fields and particles at the microlevel leads to infinities in quantum field theory, the elimination of which requires a special regularization procedure. The interaction of complicated quantum system is described by the perturbation theory, which can lead to accurate results as long as the coupling constant (quantum electrodynamics (QED)) is very small, unlike strong interaction, described by quantum chromodynamics (QCD), where the coupling constant becomes large at large distances. This problem was resolved by the the renormalization equations \cite{1.,2.}, the essence of which was  that with an increase in pulses, the coupling constant change. The study of the behavior of the strong interaction constant is associated with a large number of experimental studies \cite{3.,4.,5.}, one of which is the determination of the strong coupling constant from transverse energy–energy correlations in multĳet events at $\sqrt{s}=$13 TeV with the ATLAS detector, \cite{6.}.
	
	The aim of our article is to study the nature of the strong interaction in QCD using the renormalization group formalism, computer modeling and the latest experimental data.
	
\section{Modeling the behavior of the strong interaction constant}
This change is characterized by the formalism of the beta function connected with a so–called running coupling constant, which is standard behavior in any quantum field theory, described by the equation known as the Callan–Symanzik equation:
\[\beta(g(\mu))=\mu\frac{\partial g(\mu)}{\partial\mu},\]
which is governing the behaviour of coupling constant $g$ upon the energy scale and takes different signs at fixed points. 
	
	a) Case $\beta(g(\mu))>0$. 
	Interesting in the behavior of the coupling constants is the Landau pole - the  energy scale at which the coupling constant of a quantum field theory becomes infinite. This behavior is associated with the following type of beta function in QED
\[\beta(g)=g^3/12\pi^2, \mu\rightarrow\ +\infty,\]
which has the following dependence of the charge on the energy scale
\[e^2(\mu)=\frac{e^2(\mu_0)}{1-\frac{e^2(\mu_0)}{6\pi^2}log(\frac{\mu}{\mu_0})}\]
and singularity at the point
\[\mu=\mu_0exp[6\pi^2e^{-2}(\mu_0)].\]
In QCD the dependence of beta function and coupling pole are presented by the corresponding expressions
\[\beta(g)=3g^2/16\pi^2,\]
\[g(\mu)=\frac{g_0}{1-\frac{3}{16\pi^2}g_0log(\frac{\mu}{\mu_0})},\]
and singularity at the point
\[\mu=\mu_0exp[16\pi^2/(3g_0)].\]
Graphically, the dependence of the beta function on the coupling constants, $g$ can be represented in Figure 1.
\begin{figure}[htbp]
\bec
{\includegraphics[width=0.41\textwidth]{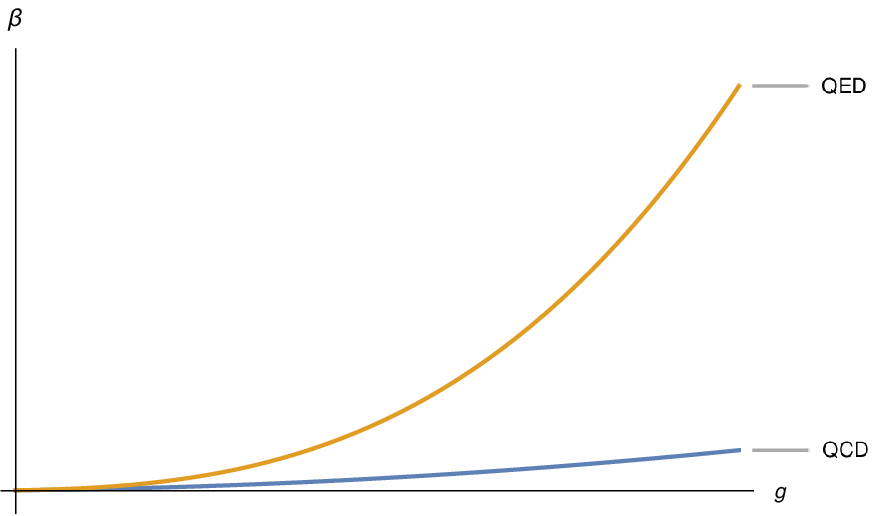}}\\
\emph{{Fig.1.}} {\emph{Dependence of the beta function on the coupling constants for $\beta(g(\mu))>0$.}}\\
\ec
\end{figure}

	b) Case $\beta(g(\mu))<0$.
The consideration of Yang-Mills theory with $\beta$ function at one-loop level
\[\beta(g)=-\frac{g^3}{8\pi^2}\Biggl(\frac{11}{2}-\frac{n_f}{3}\Biggr)\]
exhibits the asymptotic freedom 
\begin{equation}
\alpha_s(\mu)=\frac{12\pi}{(33-2n_f)log(\mu/\Lambda)}
\end{equation}
as far as the number of quark flavours is $n_f<$ 17. 
The dependence of the beta function on the coupling constants, $g$ can be represented in Figure 2.
\begin{figure}[htbp]
\bec
{\includegraphics[width=0.41\textwidth]{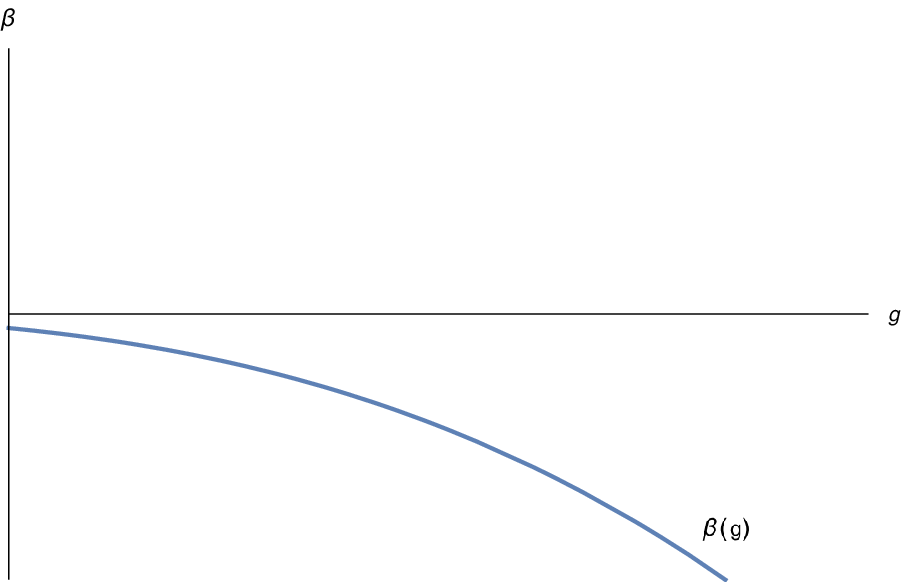}}\\
\emph{{Fig.2.}} {\emph{Schematic dependence of the beta function on the coupling constant for $\beta(g(\mu))<0$.}}\\
\ec
\end{figure}
	
	An ideal testing for perturbative Quantum Chromodynamics (pQCD) in multĳet final states was provided a the LHC in proton–proton collisions at a centre-of-mass energy of $\sqrt{s}=$13 TeV with large momentum transfer, \cite{6.}. The values of $\alpha_s(m_Z)$ are measured according to energy scale using the renormalisation group equations (RGE). The asymptotic behaviour of QCD coupling constant by comparing the measured points with the prediction of the RGE is presented in Figure 3.
\begin{figure}[htbp]
\bec
{\includegraphics[width=0.55\textwidth]{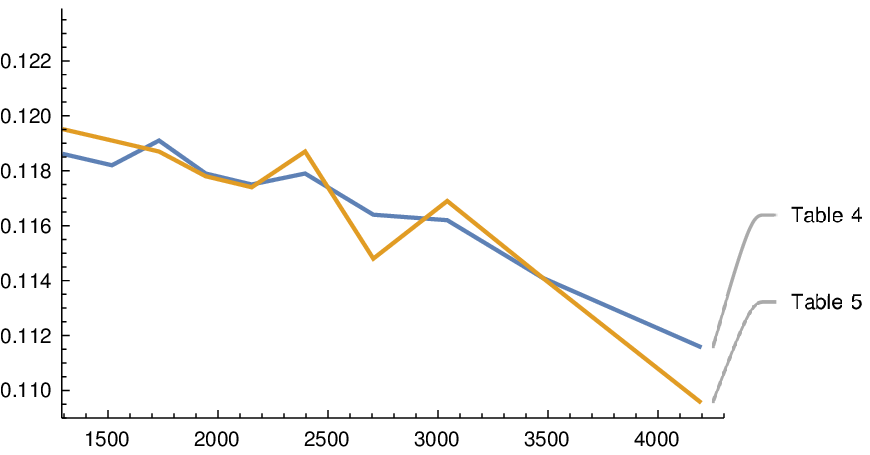}}\\
\emph{{Fig.3.}} {\emph{The values of $\alpha_s(m_Z)$ evolved to the corresponding energy scale using the RGE, data from \cite{6.}.}}\\
\ec
\end{figure}
As was emphasized in experimental paper \cite{6.},the agreement between data for the strong coupling constant 
$\alpha_s$ and theory with the renormalisation group equation is good. But  
the behavior of the constant $\alpha_s$ of equation (1) differs significantly from the nature of the behavior shown in Figure. 3
So, using Mathematica 12.0 package \cite{7.} we decided to conduct a polynomial interpolation of the nature of the behavior of the data in Tables 4 and 5 from \cite{6.} and obtained the following results for Table 4:
\[17.6798 - 0.0609999 x + 0.0000892435 x^2 - 7.1005*10^{-8} x^3 \]
\[+ 3.27592*10^{-11} x^4 - 8.35396*10^{-15} x^5 + 8.10684*10^{-19} x^6\]
\[ + 1.16153*10^{-22} x^7 - 3.68938*10^{-26} x^8 + 2.68988*10^{-30} x^9\]
 The graphical representation of the obtained data has the following character, shown in fig. 4
 \begin{figure}[htbp]
\bec
{\includegraphics[width=0.55\textwidth]{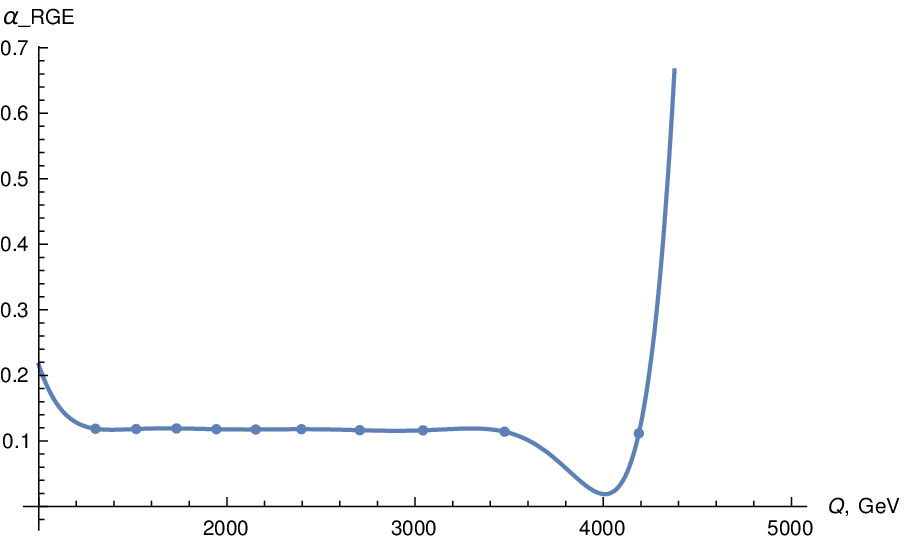}}\\
\emph{{Fig.4.}} {\emph{The values of $\alpha_s(m_Z)_{RGE}$ evolved acording to polynomial interpolation.}}\\
\ec
\end{figure}
 The nature of curve (1) is clearly pronounced only in the interval 3800-4000 GeV, Fig. 5
 \begin{figure}[htbp]
\bec
{\includegraphics[width=0.55\textwidth]{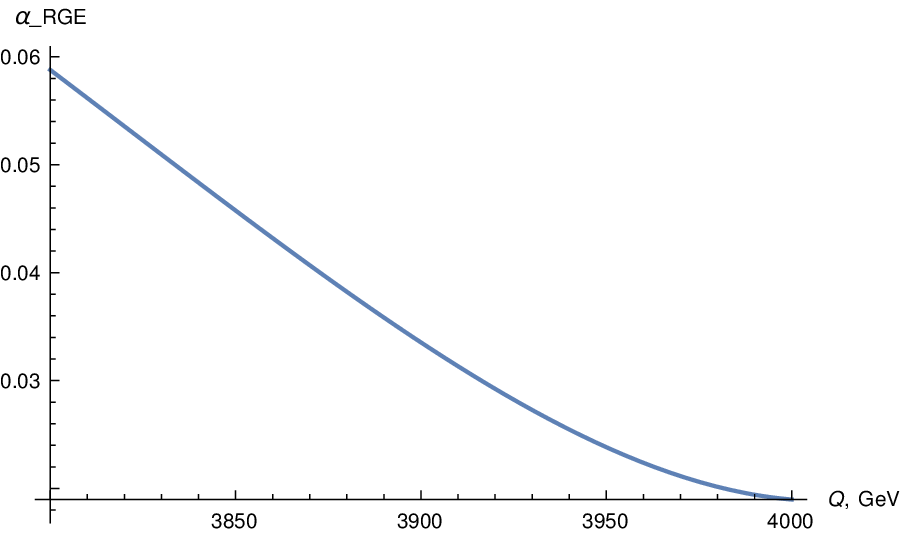}}\\
\emph{{Fig.5.}} {\emph{The values of $\alpha_s(m_Z)_{RGE}$ evolved in energy range 3800-4000 GeV.}}\\
\ec
\end{figure}
\section{Conclusion}
We have studied the nature of the behavior of the renormalization group interaction constant as a function of energy. The data obtained were used to describe the good agreement between the experimental and theoretical data in demonstrating the asymptotic freedom of QCD. We have constructed the dependence of the RGE constant on a certain energy interval and carried out a polynomial interpolation of this dependence in order to obtain information on the analytical function describing the RGE constant. It is graphically shown that good agreement with theoretical predictions regarding asymptotic freedom at high energies in QCD is observed only in the energy region 3800-4000 GeV.

\end{document}